\documentclass{article}

\pdfoutput=1 
\usepackage{graphicx}

\usepackage[numbers]{natbib}

\begin{document}


\title{Proton Colliders at the Energy Frontier}
\author{Michael Benedikt and Frank Zimmermann \protect\\
CERN, CH-1211 Geneva 23, Switzerland}

\maketitle

\begin{abstract}
Since the first proton collisions at the CERN Intersecting Storage Rings (ISR)
\protect\cite{isr,isr-myers}, 
 hadron colliders have defined the energy frontier \cite{scandale-rast}.  
Noteworthy are the conversion of the 
Super Proton Synchrotron (SPS) \cite{sps,lyn-sps}
into a proton-antiproton collider, the Tevatron 
proton-antiproton collider \cite{tevatron},
as well as the abandoned SSC in the United States \cite{ssc,wienands-ssc}, 
and early forward-looking studies of even 
higher-energy colliders \cite{keil-100tev,keil-SR,barletta-sc-sd,vlhc}.    
Hadron colliders are likely to determine the pace of 
particle-physics progress also during the next hundred years.
Discoveries at past hadron colliders were essential for 
establishing the so-called Standard Model of particle physics. 
The world's present flagship collider, the Large Hadron Collider (LHC)
\cite{lhc}, including its high-luminosity 
upgrade (HL-LHC) \cite{hl-lhc}, 
is set to operate through the second half of the 2030's. 
Further increases of the energy reach during the 21st century require another, still 
more powerful hadron collider.
Three options for a next hadron collider are presently under investigation. 
The Future Circular Collider (FCC) study, hosted by CERN, 
is designing a 100 TeV collider, to be installed 
inside a new 100 km tunnel in the Lake Geneva basin. 
A similar 100-km collider, called Super proton-proton Collider (SppC), 
is being pursued by CAS-IHEP in China. 
In either machine, for the first time in hadron storage rings,
synchrotron radiation damping will be significant, 
with a damping time of the order of 1 hour.
In parallel, the synchrotron-radiation power emitted inside the cold magnets becomes an important design constraint.
One important difference between FCC and SppC is the magnet technology. 
FCC uses 16 Tesla magnets based on Nb$_3$Sn superconductor, while SppC magnets shall be realized with 
cables made from iron-based high-temperature superconductor.
Initially the SppC magnets are assumed to provide a more
moderate dipole field of 12 T, but they can later 
be pushed to a final ultimate field of 24 T.
A third collider presently under study is the High-Energy LHC (HE-LHC), 
which is a higher energy collider in the existing LHC tunnel, 
exploiting the FCC magnet technology in order to essentially double the LHC energy at significantly higher luminosity.
\end{abstract}


\section{Introduction}

Circular hadron colliders are known as discovery machines. Their discovery reach is determined by the beam energy, which 
depends on only two parameters: the dipole magnetic field and the size of the collider. Therefore, historically new 
colliders always were larger and used stronger magnets than their predecessors. For example, the Tevatron near Chicago
was the first hadron collider based on superconducting magnet technology, 
with a dipole field of 4.2 T, and it was installed in a 6.3 km ring. 
The LHC uses 8.3 T dipoles in a 26.7 km tunnel. The 100 TeV Future Circular Collider (hadron version ``FCC-hh'') requires 16 Tesla dipole 
magnets in a 100 km ring. No other proposed concept, 
not even a muon collider or a plasma collider, 
appears technically ready to provide 
collision energies in the 10's of 
TeV energy range during the 21st century.

\begin{figure}[htbp]
\centering
\includegraphics[width=0.95\columnwidth]{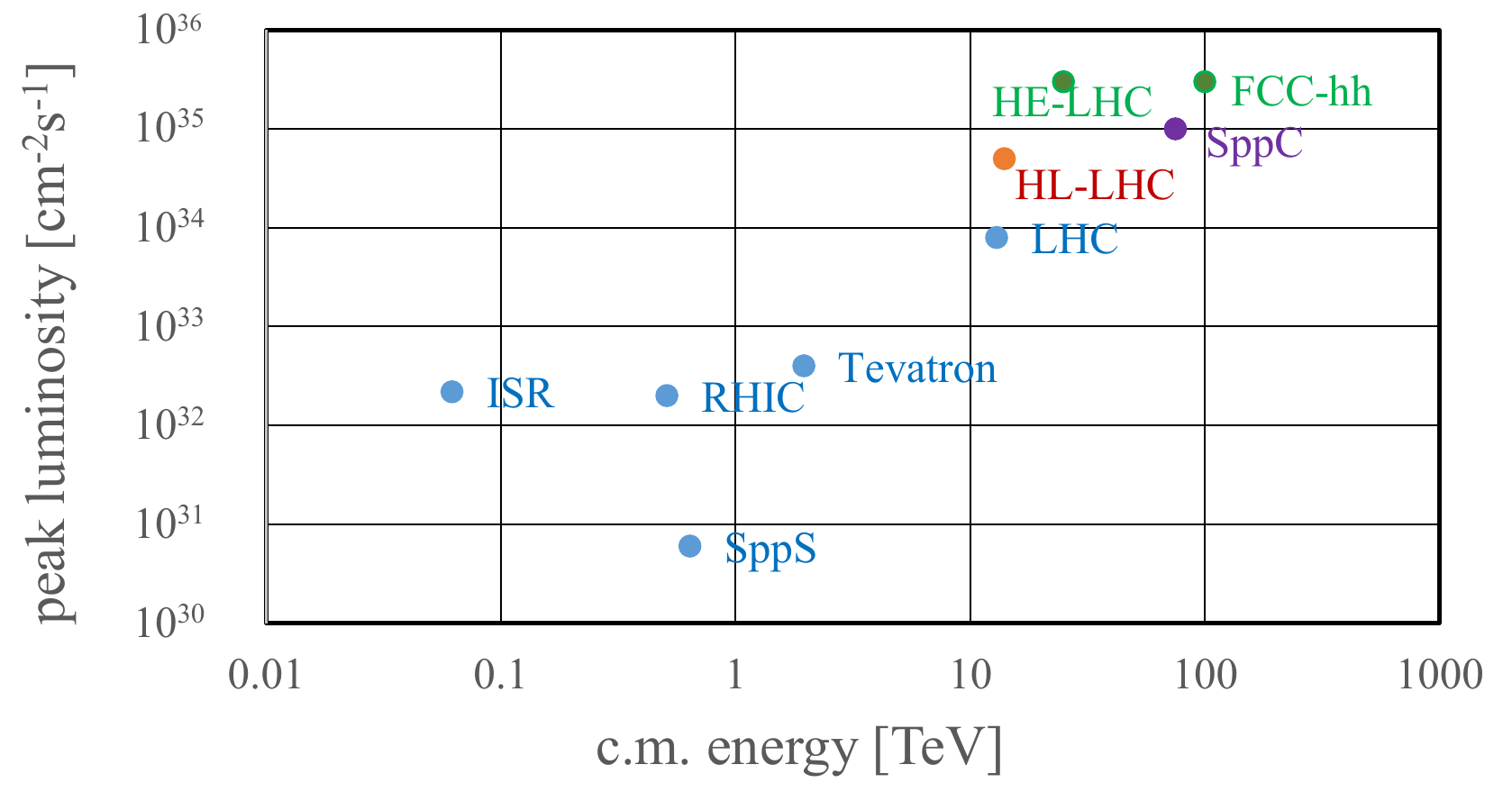}
\caption{Luminosity vs.~centre-of-mass energy for past and present [blue], upcoming [red], and longer-term future hadron 
($pp$ or $p\bar{p}$) colliders [green and purple] around the world.}
\label{coll-hist}
\end{figure}

\begin{figure}[htbp]
\centering
\includegraphics[width=0.98\columnwidth]{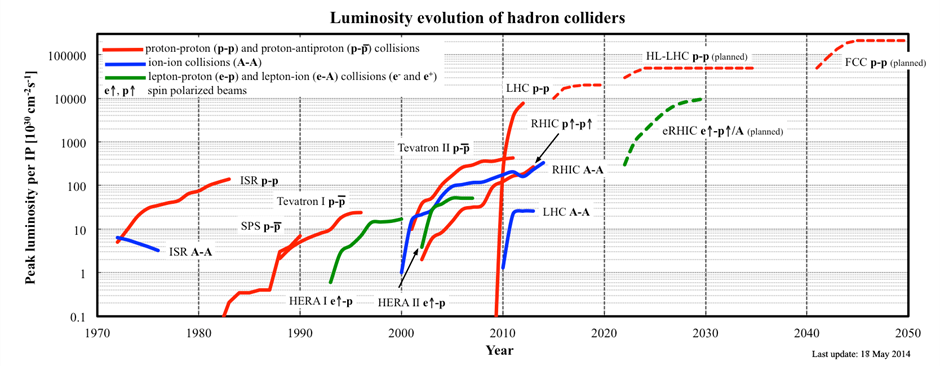}
\caption{Hadron collider peak luminosity as a function of year for 
past, operating, and proposed facilities including the Future Circular Collider (courtesy W.~Fischer).}
\label{coll-hist2}
\end{figure}

The collider luminosity ideally increases with the square of the energy since the cross sections decrease as the inverse square  of energy. 
However, due to the nonlinear parton distribution inside the colliding protons also a lower luminosity can produce exciting
 physics, and the most important parameter of a hadron collider remains its energy. 
Nevertheless, at a given energy the discovery reach grows with higher luminosity \cite{salam}. which is one of 
the motivations for upgrading the LHC to the HL-LHC. 
The LHC design has already dramatically 
increased the luminosity compared with previous machines. This can be seen in 
Figs.~\ref{coll-hist} and \ref{coll-hist2}. Much higher luminosities still are expected for the approved HL-LHC, 
which will lower its peak luminosity by ``levelling''  in order to make it acceptable for the physics experiments,  
as well as for the proposed HE-LHC and FCC-hh. The luminosity for the latter two machines will profit from significant 
radiation damping at the associated high beam energies and magnetic fields \cite{mbdsfz}.

\section{Hadron-Collider Beam Dynamics and Limitations}
The hadron-collider luminosity increases linearly 
with energy due to the shrinking beam sizes, when 
keeping the beam current, the beta functions at the interaction point (IP), 
$\beta_{x,y}^{\ast}$, and the beam-beam tune shift constant.   
Even higher luminosity can be achieved by reducing the IP beta functions. 
Perhaps surprisingly, until now all hadron 
colliders, starting from the ISR, have operated with similar beta functions, 
with minimum values of about 0.3 m; see Table \ref{tab:beta}.
With 0.15~m (or even 0.10~m) the HL-LHC will set a new record. An ongoing study aims at pushing the FCC-hh $\beta^{\ast}$ down to 5 cm 
\cite{martin}. 

For proton-proton colliders with many bunches, such as HL-LHC and FCC-hh, 
a crossing angle 
is required to avoid or mitigate parasitic beam-beam collisions. Unfortunately, this crossing angle needs to be increased as $\beta_{x,y}^{\ast}$ is reduced. 
Without countermeasures 
this would dramatically degrade the geometric overlap of the colliding bunches and all but eliminate any benefit 
from reducing the IP beam size. To avoid this degradation, 
the HL-LHC, the HE-LHC and FCC-hh phase 2 will all use novel crab cavities 
\cite{crab1,crab2,crab3,crab4,crab5,crab6}. 
These are transversely deflecting RF cavities, which impart kicks of opposite sign tothe head and the tail of each bunch, so as to 
maintain the crossing angle of the bunch centroid motion, while 
at the same time restoring the full geometric 
bunch overlap during the collision. 

\begin{table}[htbp]
\caption{Beta* at hadron colliders (R.~Tomas \protect\cite{xbeam17}).}
\label{tab:beta}
\begin{center}
\begin{tabular}{|l|c|c|}
\hline
collider & $\beta_{x}^{\ast}$ [m] & $\beta_{y}^{\ast}$ [m] \\ 
\hline 
ISR & 3.0 & 0.3 \\
\hline
S$p\bar{p}$S & 0.6 & 0.15 \\
\hline
HERA-p & 2.45 & 0.18\\
\hline
RHIC & 0.50 & 0.50 \\
\hline
Tevatron & 0.28 & 0.28 \\
\hline
LHC & 0.3 & 0.3 \\
\hline
HL-LC& 0.15 & 0.15 \\ 
\hline
FCC-hh & 1.1$\rightarrow$ 0.3 (0.05) & 1.1$\rightarrow$ 0.3 (0.05)\\
\hline
SppC & 0.71 & 0.71  \\
\hline
HE-LHC & 0.25 & 0.25 \\
\hline
\end{tabular}
\end{center}
\end{table}

Present and future hadron colliders are characterized by a large amount of stored beam energy, which render machine protection 
a paramount concern. A multi-stage collimation system is needed to avoid local beam loss spikes near cold magnets, which would
 induce magnet quenches. The collimation system of the LHC 
\cite{lhc-coll} 
works according to specification \cite{lhc-coll2}. 
For the planned and proposed 
future colliders --- HL-LHC, HE-LHC, SppC, and FCC-hh --- collimation remains a challenge.

Beam injection and beam extraction are particularly sensitive operations, as the injection or dump kickers belong to the fastest
 elements in the machine. The collider design must be robust against the sudden asynchronous firing of a kicker unit. 
The collimators are likely to be the first element to be hit by the beam in case of any fast failure. They must withstand
 the impact of one or a few bunches. 
 The primary and secondary 
 collimators of the LHC are based on carbon-carbon composite material.  
 For the HL-LHC and farther future machines, even stronger materials   
 are being developed and examined, which, in addition, 
feature a higher conductivity and, hence, lower ``impedance''. 
More advanced options include the use of short 
bent crystals as primary collimators, and the deployment of hollow electron-beam lenses as non-destructible collimators.
An acceptable performance of the collimation system along with small IP beta function also requires an excellent optics control.

Hadron beam intensity may be limited by conventional instabilities, in particular,  due to the very large circumference and low
 momentum compaction, respectively, by resistive wall instability (low revolution frequency) and transverse mode coupling at injection. 
Another intensity limit may arise from the build up of an electron cloud, 
which may drive a different type of instability or create additional 
significant heat loads on the beam screen inside the cold magnets. 
Indeed, the electron cloud is a primary source of
 beam instability already 
in the LHC, especially with a proton bunch spacing of 25 ns. 
The beam performance tends to improve in 
time thanks to beam-induced surface conditioning (``scrubbing''). In addition, at the LHC occasional losses of transverse 
or longitudinal Landau damping arise due to the classical machine impedance with contributions from the resistive vacuum chamber, RF cavities, and chamber transitions. Concerning instability mitigation, the following lessons 
have been learnt \cite{emetral1,emetral2,emetral3} in operating 
the LHC \cite{xbeam17}: There exists a narrow range of machine settings for which the beam remains stable all along the cycle; 
instabilities occur if the transverse betatron coupling exceeds a certain threshold value (different at different
 stages of operation); chromaticity settings are crucial along the cycle and cannot be relaxed; the octupole-magnet settings have
 to be adapted according to beam emittance; finally, the transverse damper is indispensable to preserve beam stability all along the cycle.

\begin{figure}[htbp]
\centering
\includegraphics[width=0.95\columnwidth]{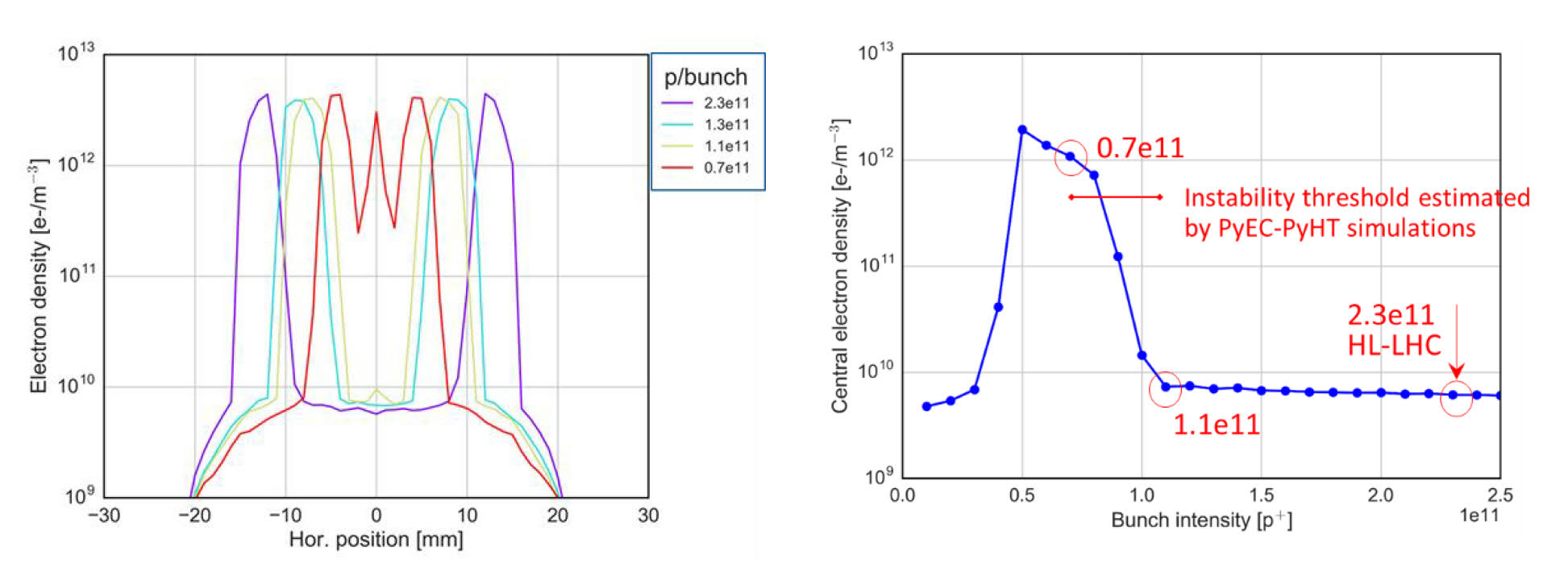}
\caption{Electron cloud distribution in an LHC dipole 
magnet for varying proton bunch population (left) and central electron density versus bunch intensity (right) (A.~Romano 
et al.; and G.~Rumolo, Valencia, 2017 \protect\cite{xbeam17}).}
\label{ecloud}
\end{figure}

Interestingly, the electron cloud can drive coherent instabilities even when beams are in collision, with the associated strong Landau damping. Simulations and earlier measurements at the 
SPS show that, for lower
 bunch intensities, the electron cloud in the dipoles tends to form a central stripe; at the LHC the central 
density threshold of the electron-cloud driven single-bunch head-tail instability
 ($\sim 5\times 10^{11}$~m$^{-3}$ at a chromaticity of  
$Q'\approx 15$) is crossed when the bunch intensity decreases, as 
is illustrated in Fig.~\ref{ecloud}; for $Q'>20$ the threshold becomes much higher. This 
explanation of the beam instabilities observed towards the end of LHC physics fills also is 
consistent with the disappearance of the phenomenon after scrubbing.

\section{Pushing the Energy Frontier in the 21st Century}
A very large circular hadron collider with 100 TeV c.m.~collision energy will offer 
access to new particles through direct production  
in the few-TeV to 30 TeV mass range, far beyond the LHC reach \cite{Mangano:2270978}.
It would also provide 
much-increased rates 
for phenomena in the sub-TeV mass range and,
thereby, a much increased precision compared with the HL-LHC \cite{Mangano:2270978}.  

The centre-of-mass 
energy reach of a hadron collider is directly proportional to the maximum magnetic field $B$ 
and to the bending radius $\rho$:  
\begin{equation}
E_{\rm c.m.} \propto \rho  B .
\end{equation}
Therefore, an increase in the size of the collider compared with the 
LHC by a factor of about 4 and an approximate doubling of the magnetic 
field yields almost an order of magnitude increase in energy.

Such approach was first suggested during the High-Energy LHC (HE-LHC) workshop in 2010 \cite{Todesco:1344820}. 
Now it is the focus of the Future Circular Collider (FCC) study \cite{benedikt}, 
which was launched in response to the 2013 Update of the 
European Strategy for Particle Physics \cite{strategy2013}:
Dipole magnets with a field of 16 Tesla in a 100 km ring will result in a centre-of-mass energy of 100 TeV. 
This goal defines the overall infrastructure requirements for the FCC accelerator complex.
The FCC study scope also includes the design of a high-luminosity 
e$^+$e$^-$ collider (FCC-ee) operating at c.m.~energies of 90--365 GeV, 
as a possible first step  --- with a remarkably rich 
physics programme \cite{tlepphysics} ---,  
as well as a proton-electron collision option (FCC-he) at one interaction point, 
where a 60 GeV electron beam from an energy recovery linac 
would be collided with one of the two 50 TeV proton beams circulating in the FCC-hh.
The design of a higher-energy hadron collider in the LHC tunnel 
based on FCC-hh magnet technology 
--- the so-called High-Energy LHC (HE-LHC) 
--- is yet another part of the FCC study.

As of December 2017, 113 institutes and 25 companies from 32 countries are participating in the FCC study.
The near-term goal is to deliver a Conceptual Design Report of 
all FCC collider options, including technologies, detector design, and physics goals,
before the end of 2018, as input to the next European
Strategy Update process planned for 2019/20.

\begin{figure}[htbp]
\centering
\includegraphics[width=0.9\columnwidth]{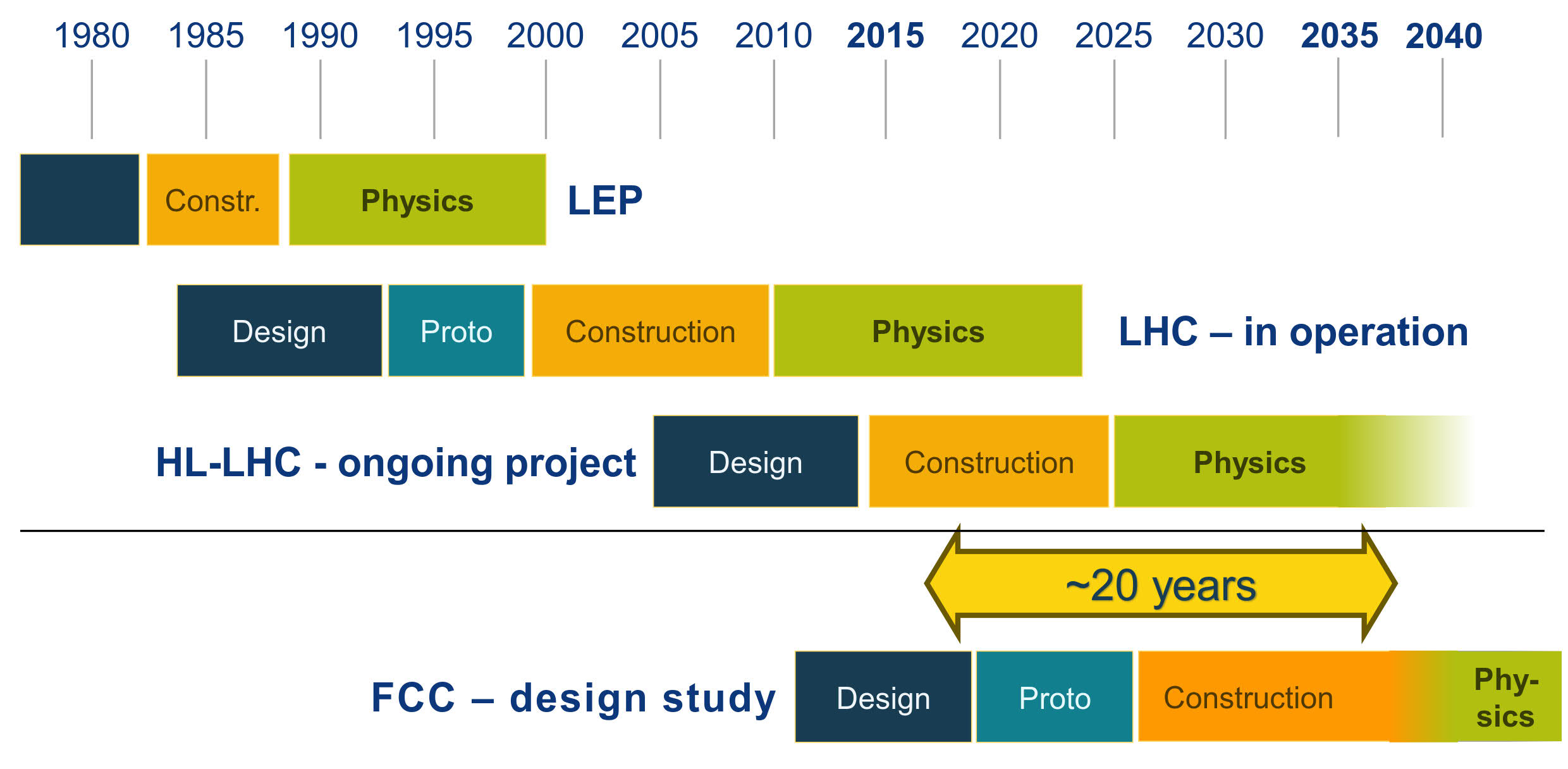}
\caption{Time lines of several 
past, present and future circular colliders at CERN, distinguishing periods of design, prototyping, construction, and physics exploitation.}
\label{fig2}
\end{figure}

Figure \ref{fig2} compares the time lines of various past 
and present circular colliders at CERN with a projected time line for the FCC, indicating a need for fast progress.

CEPC and SppC are two colliders similar to FCC-ee/FCC-hh,
which are being studied by another international collaboration, centred at IHEP Beijing \cite{accel:cepc}. 
These two machines have a similar circumference as FCC, 
of about 100 km. 
Several possible locations in China are under study.  
The e$^+$e$^-$ collider CEPC is designed with a maximum centre-of-mass energy of 240 GeV, and noticeably 
lower luminosity than FCC-ee.  
The SppC hadron collider relies on 12 T (later 24 T) iron-based high-temperature superconducting magnets, 
which could be installed in the same tunnel as the CEPC.  

Table \ref{tabhadron} shows key parameters of FCC-hh, SppC, and HE-LHC, together  
with the design values of the present LHC and its luminosity upgrade (HL-LHC).

\begin{table}[htbp]
\begin{center}
\begin{tabular}{|l|c|c|c|c|c|}
\hline
parameter & 
\multicolumn{2}{|c|}{FCC-hh} & 
SppC & 
HE-LHC
& (HL-)LHC \\
\hline 
c.m.~energy [TeV] 
& \multicolumn{2}{|c|}{100} 
& 75 
& 27 & 14 \\
dipole field  [T] 
& \multicolumn{2}{|c|}{16} 
& 12  
& 16 & 8.3 \\
\hline 
circumference [km] 
  & \multicolumn{2}{|c|}{97.8} 
& 100  
& 26.7 & 26.7 \\
\hline 
beam current [A] 
  & \multicolumn{2}{|c|}{0.5} 
& 0.77   
& 1.12 & (1.12) 0.58  \\
part./bunch [$10^{11}$] 
  & \multicolumn{2}{|c|}{1} 
& 1.5   
& 2.2 & (2.2) 1.15  \\
bunch spacing [ns]  
  & \multicolumn{2}{|c|}{25} 
& 25 
& 25 & 25 \\
norm.~emittance $\varepsilon_{N}$ [$\mu$m]  
  & \multicolumn{2}{|c|}{2.2 (1.1)}
& 3.16  
& 2.5 (1.25) & (2.5) 3.75 \\
\hline 
IP beta function [m]  
  & 1.1 & 0.3 
& 0.71   
& 0.25 & (0.15) 0.55 \\
lum.~[$10^{34}$ cm$^{-2}$s$^{-1}$]  
  & 5 & 30  
& 10   
& 28 & (5, lev.) 1 \\
events per crossing
  & 170 & 1000  
& $\sim$300   
& 800 & (135) 27 \\
\hline 
SR power/beam [kW]
& \multicolumn{2}{|c|}{2400} 
& 1130   
& 100 & (7.3) 3.6 \\
longit.~damp.~time [h]
& \multicolumn{2}{|c|}{1.1} 
& 2.4  
& 3.6 & 25.8 \\
\hline
init.~burn-off time [h]
& 17 & 3.4
& 13    
& 3.0 & (15) 40 \\
\hline
\end{tabular}
\end{center}
\caption{Parameters of future hadron colliders, 
the LHC and its HL-LHC upgrade. 
The HL-LHC will level the luminosity at a value of 
$5\times 10^{34}$~cm$^{-2}$s$^{-1}$, 
for the particle-physics
experiments; its virtual peak luminosity is 
about 5 times higher.}
\label{tabhadron} 
\end{table}

\section{Beam Parameter Evolution during a Physics Fill} 
At the LHC and HL-LHC radiation damping during a physics fill is almost
 negligible. The HL-LHC requires luminosity leveling via changes in $\beta^{\ast}$ and crossing angle, in order to limit the event pile up to a value acceptable for the physics detectors \cite{yannis-cham2017}.

For future higher-energy hadron colliders radiation damping becomes significant. In such a situation, the longitudinal emittance needs to be kept constant during the physics store, through controlled longitudinal noise excitation, in order to maintain longitudinal Landau damping \cite{zimmermann2001}. 
At the same time, the transverse emittance shrinks due to the strong radiation damping, while the proton intensity rapidly decreases as the result of the high luminosity.

The initial proton burn-off time can be computed as
\begin{equation}
\tau_{\rm bu} = \frac{N_{b}n_{b}}{L_0 \sigma_{\rm tot} n_{\rm IP}}\; ,
\end{equation}
where $N_{b}$ denotes the bunch population, $L_{0}$ the initial 
luminosity, $\sigma_{\rm tot}$ the total proton-proton cross section,    
$n_{b}$ the number of bunches per beam,
and $n_{\rm IP}$ the number of high-luminosity interaction points (IPs);
$n_{\rm IP}=2$ for all four colliders under consideration.

For the FCC-hh the emittance damping time is shorter than the proton burn-off time. 
As a result the total beam-beam tune shift 
\begin{equation}
\Delta Q_{\rm bb} = n_{\rm IP} \frac{ r_{p} N_{b}}{4 \pi \varepsilon_{N}}
\end{equation} 
increases during the store ($r_{p}$ designates the 
``classical proton radius'').  
At some point the beam-beam limit is reached,
and the tranverse emittance must then be controlled by transverse 
noise excitation, so as to keep the beam-beam tuneshift at, or below, the empirical limit. 
This determines the further luminosity evolution during the store and the optimum run time 
\protect\cite{mbdsfz}. 

\begin{figure}[htp]
\centering
\includegraphics[width=0.45\columnwidth]{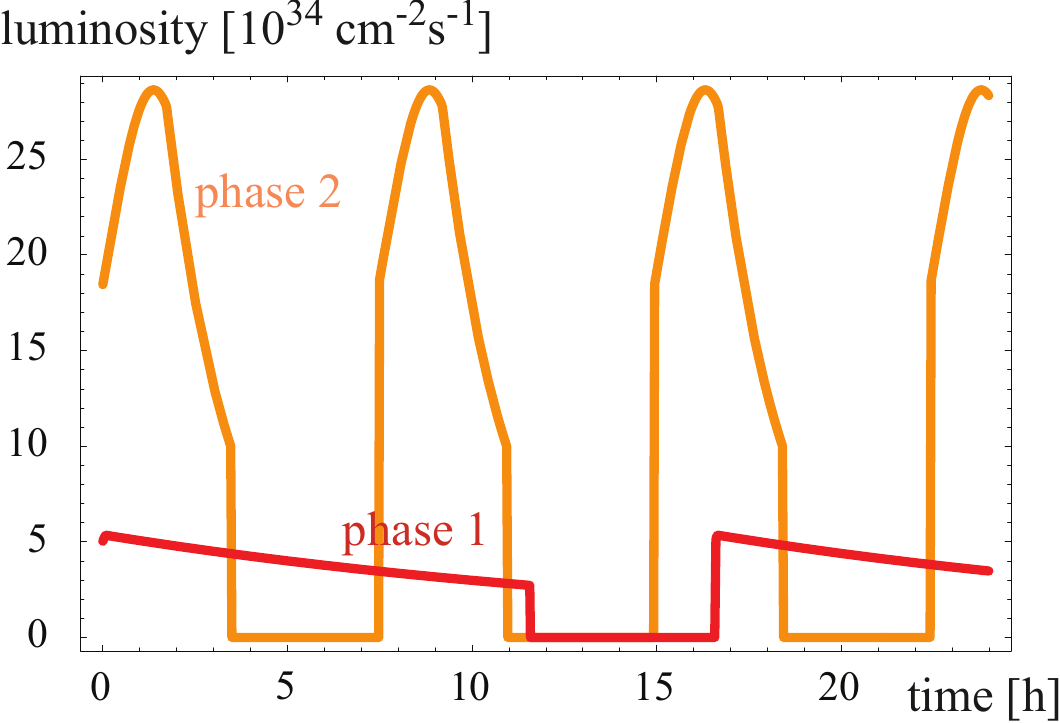}
\includegraphics[width=0.45\columnwidth]{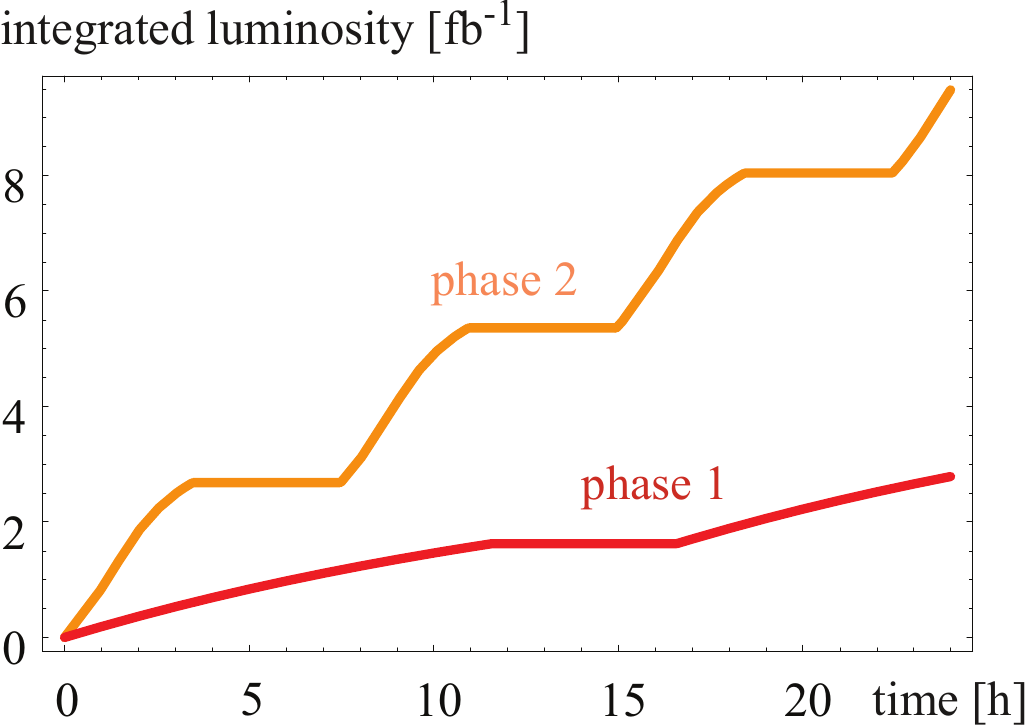}
\includegraphics[width=0.45\columnwidth]{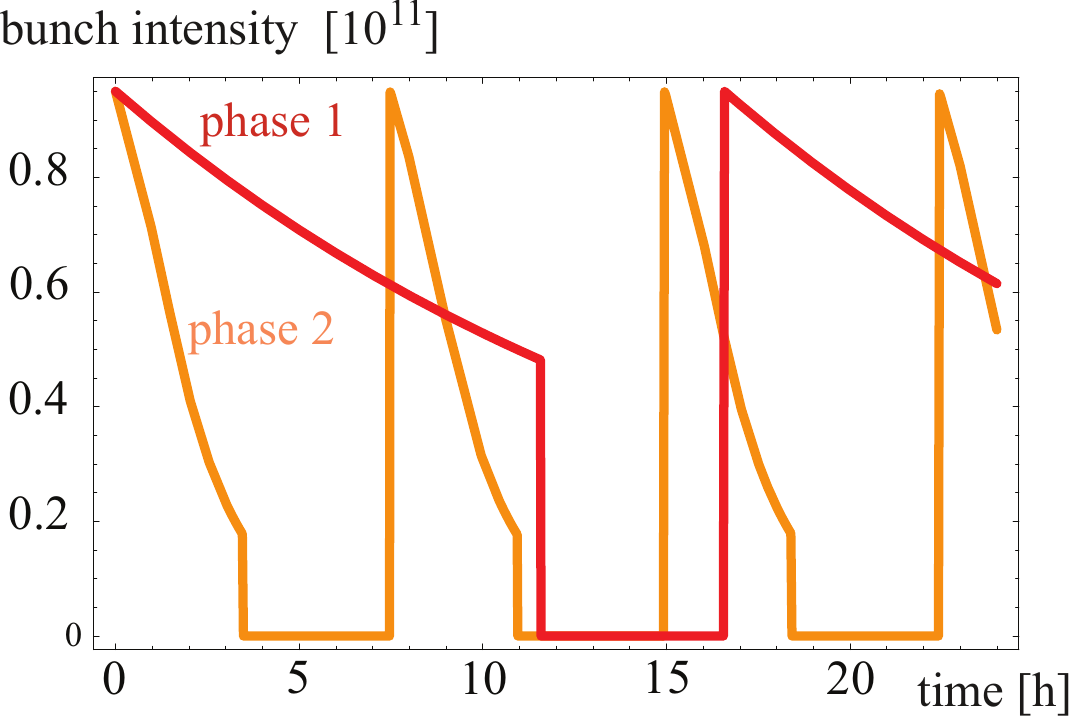}
\includegraphics[width=0.45\columnwidth]{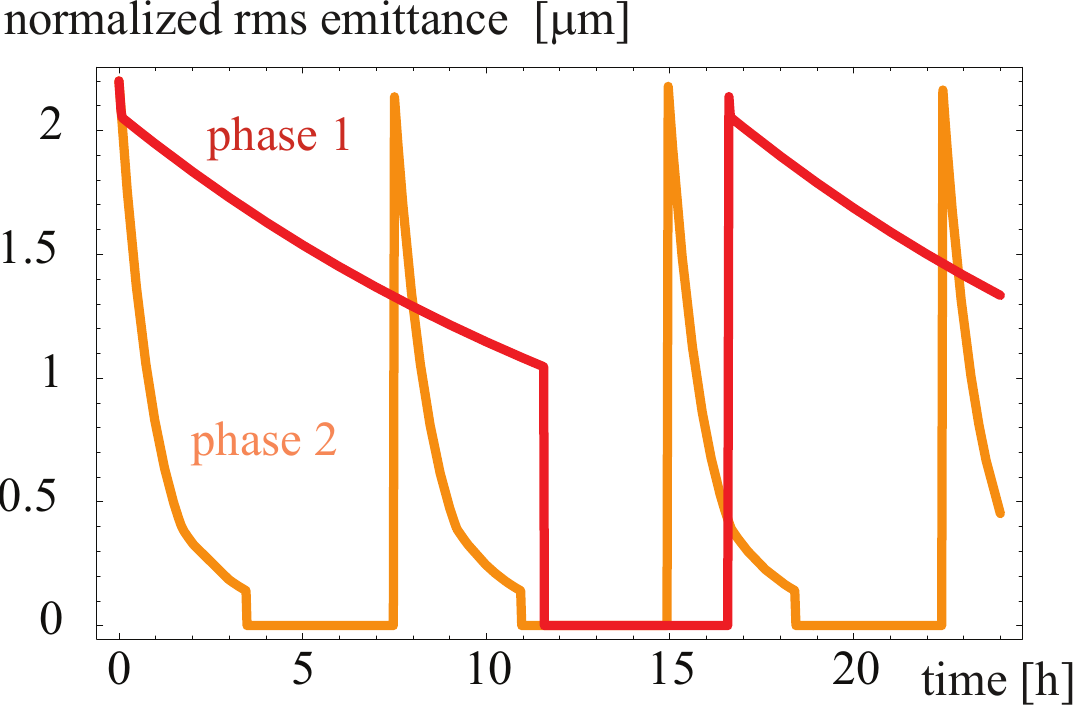}
\includegraphics[width=0.45\columnwidth]{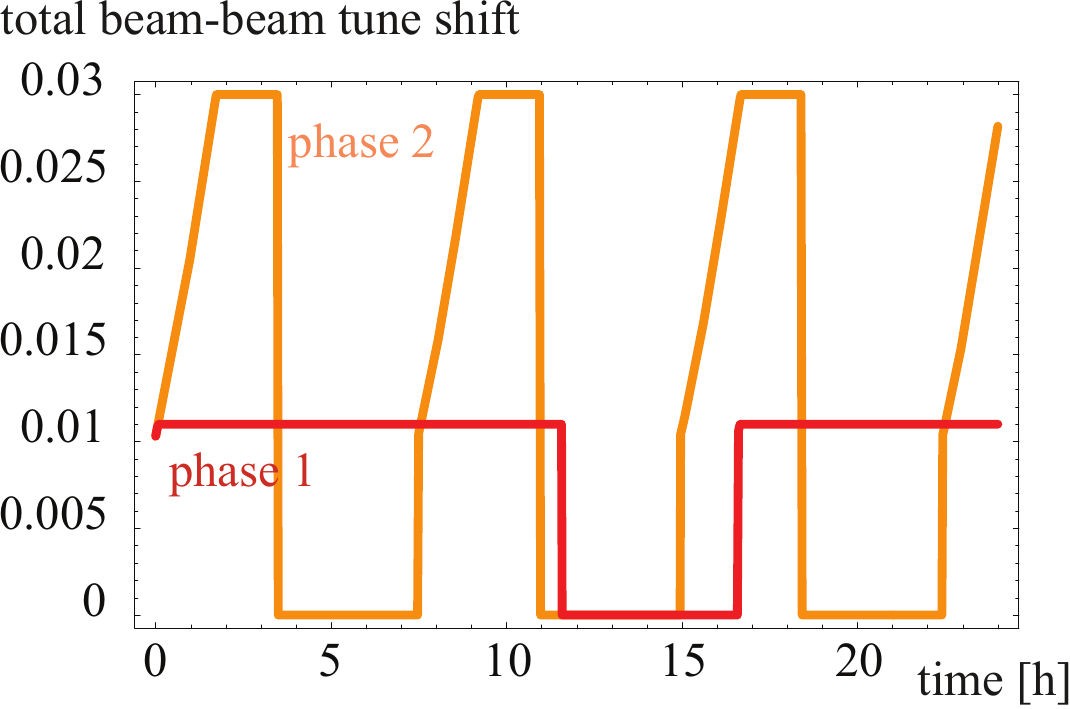}
\includegraphics[width=0.45\columnwidth]{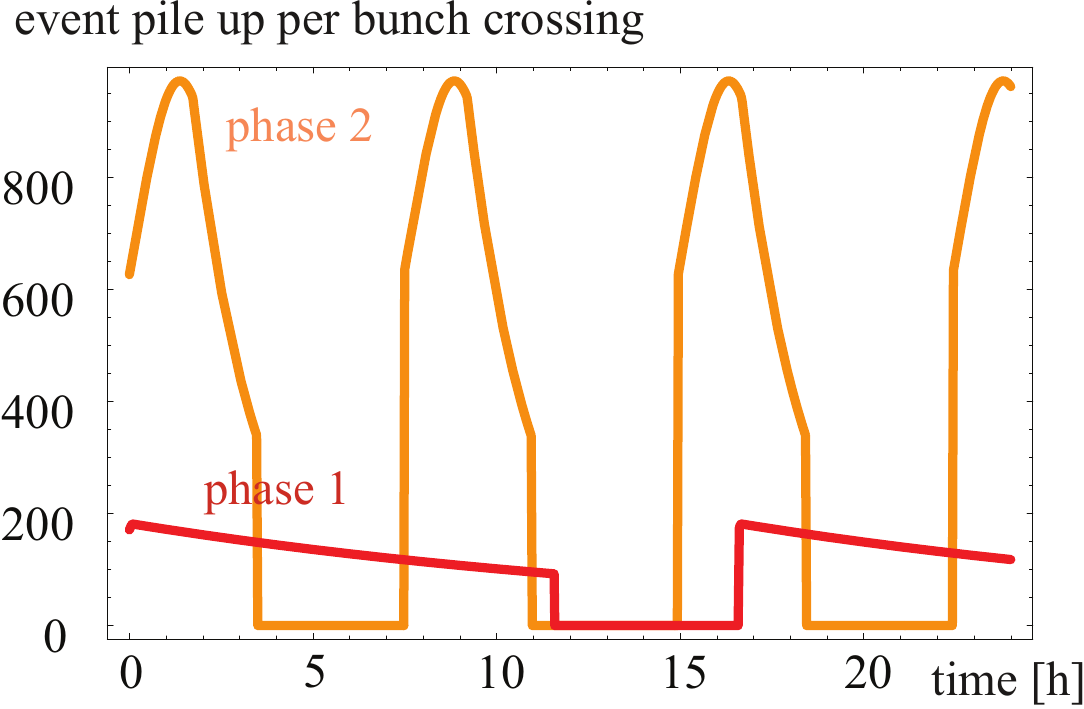}
\caption{Instantaneous luminosity, integrated luminosity, bunch population,
emittance, total beam-beam tune shift, and event pile up 
as a function of time during 24 hours with 25 ns bunch spacing, for  
FCC-hh Phases 1 ($\beta_{x,y}^{\ast}=1.1$ m, $\Delta Q_{\rm bb} =0.01$) 
and 2 ($\beta_{x,y}^{\ast}=0.3$ m, $\Delta Q_{\rm bb} =0.03$) 
\protect\cite{mbdsfz}.}
\label{fig2b}
\end{figure}

Figure \ref{fig2b} presents the evolution of instantaneous luminosity,
integrated luminosity, bunch population, emittance, pile up and beam-beam tune shift for both phases 
of FCC-hh over 24 h of running. 
Here, we assume that the injected beam corresponds to the baseline parameters  
and a total beam-beam tune shift (sum of two IPs) 
of $\Delta Q_{\rm tot}\approx 0.01$.
In Phase 2 the emittances are allowed to shrink, the tune shift increases during the fill, 
until the higher tune-shift limit of $\Delta Q_{\rm tot}= 0.03$ is reached.
From this moment onwards the further emittance damping is again 
counterbalanced by a controlled blow up keeping the beam brightness constant. 
Only the proton burn-off in collision and the natural, 
or --- after reaching the beam-beam limit ---  
controlled emittance shrinkage due to radiation damping   
are taken into account. Other additional phenomena like 
gas scattering, Touschek effect, intrabeam scattering, and 
space charge are insignificant for the 
50 TeV beams, in the scenarios considered.

By contrast,
at the HE-LHC the proton burn off time is slightly shorter than the radiation damping time.
This situation is qualitatively different from the FCC-hh. 
For the HE-LHC there is almost a natural luminosity 
leveling, while the beam-beam tune shift naturally decreases during the store.

Following a derivation similar to those in Ref.~\cite{mbdsfz},
for the HE-LHC the integrated luminosity per interaction point (IP)  
at time $t$ during the fill is 
\begin{equation}
\int_{0}^{t} L(t) dt
= \frac{f_{\rm rev} N_{b,0}^{2} n_{b}}{
4 \pi \varepsilon_{0} \beta_{x,y}^{\ast}} \; \frac{\tau}{B} \left( 1 - \frac{1}{1 - B + B \exp \left(
  t/\tau\right) } \right) \; .
\end{equation} 
The optimum run time $t_{r,{\rm opt}}$ then follows from 
\begin{equation}
\left[
(1-B) \exp \left(-t_{r}/\tau\right) +
(2 B-1) - B \exp \left(t_{r}/\tau\right)
+t_{r}/\tau+t_{\rm ta}/\tau 
\right]_{\rm t_{r}=t_{r,{\rm opt}}}
\stackrel{!}{=} 0\; ,
\end{equation}
where $t_{\rm ta}$ denotes the average turnaround time, $\varepsilon _{0}$ the initial geometric rms emittance, $N_{b,0}$ the initial bunch population,
$f_{\rm rev}$ the revolution frequency,
$n_{IP}$ the number of high-luminosity collision points, $\sigma_{\rm tot}$ the total cross section, 
$\tau$ the transverse emittance damping time,
and 
\begin{equation}
B\equiv \frac{\sigma_{\rm tot} n_{IP} f_{\rm rev}
N_{b,0} \tau  }{4 \pi \beta_{x,y}^{\ast} \varepsilon_{0} }\; .
\end{equation}

\begin{figure}[htp]
\centering
\includegraphics[width=0.45\columnwidth]{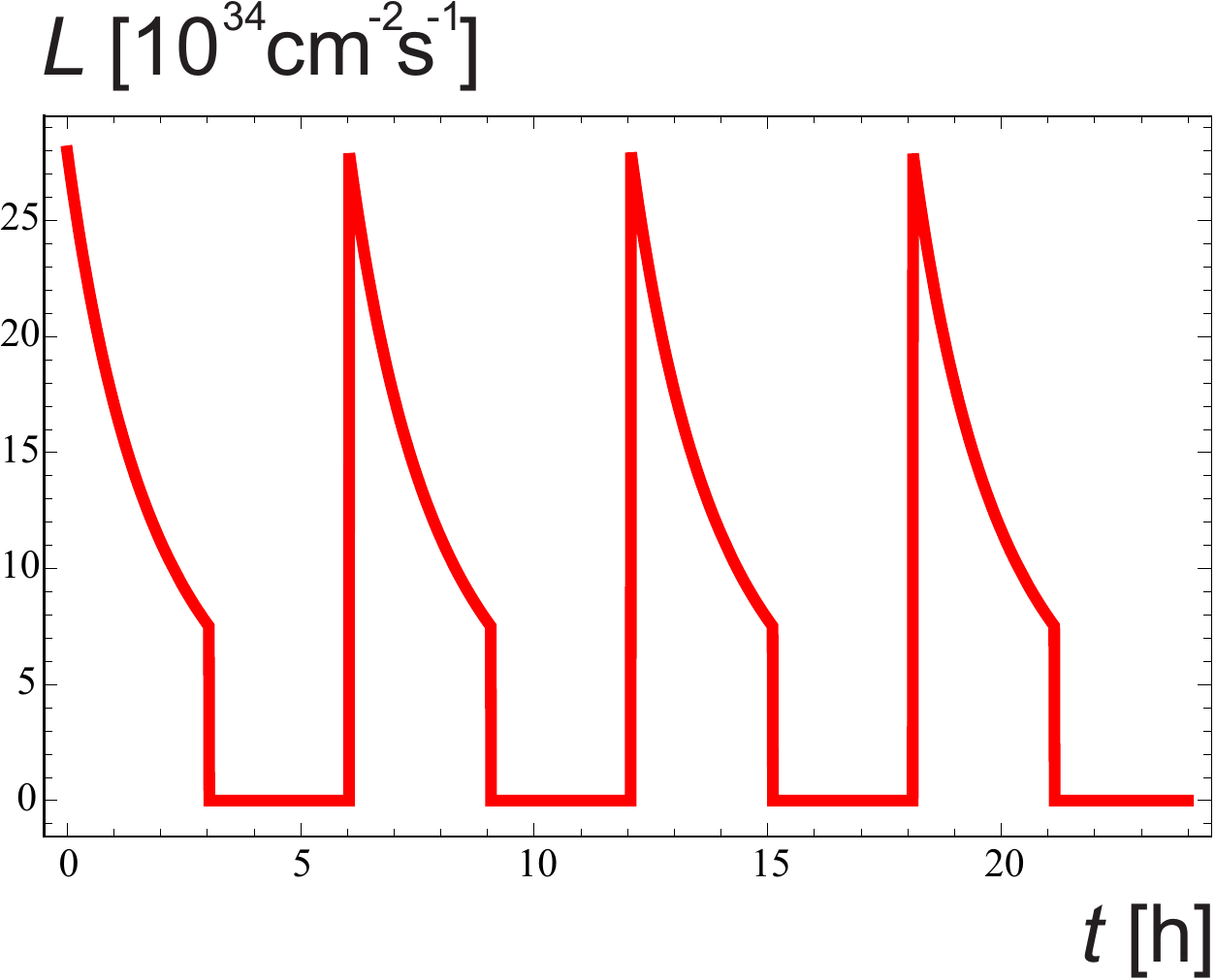}
\includegraphics[width=0.45\columnwidth]{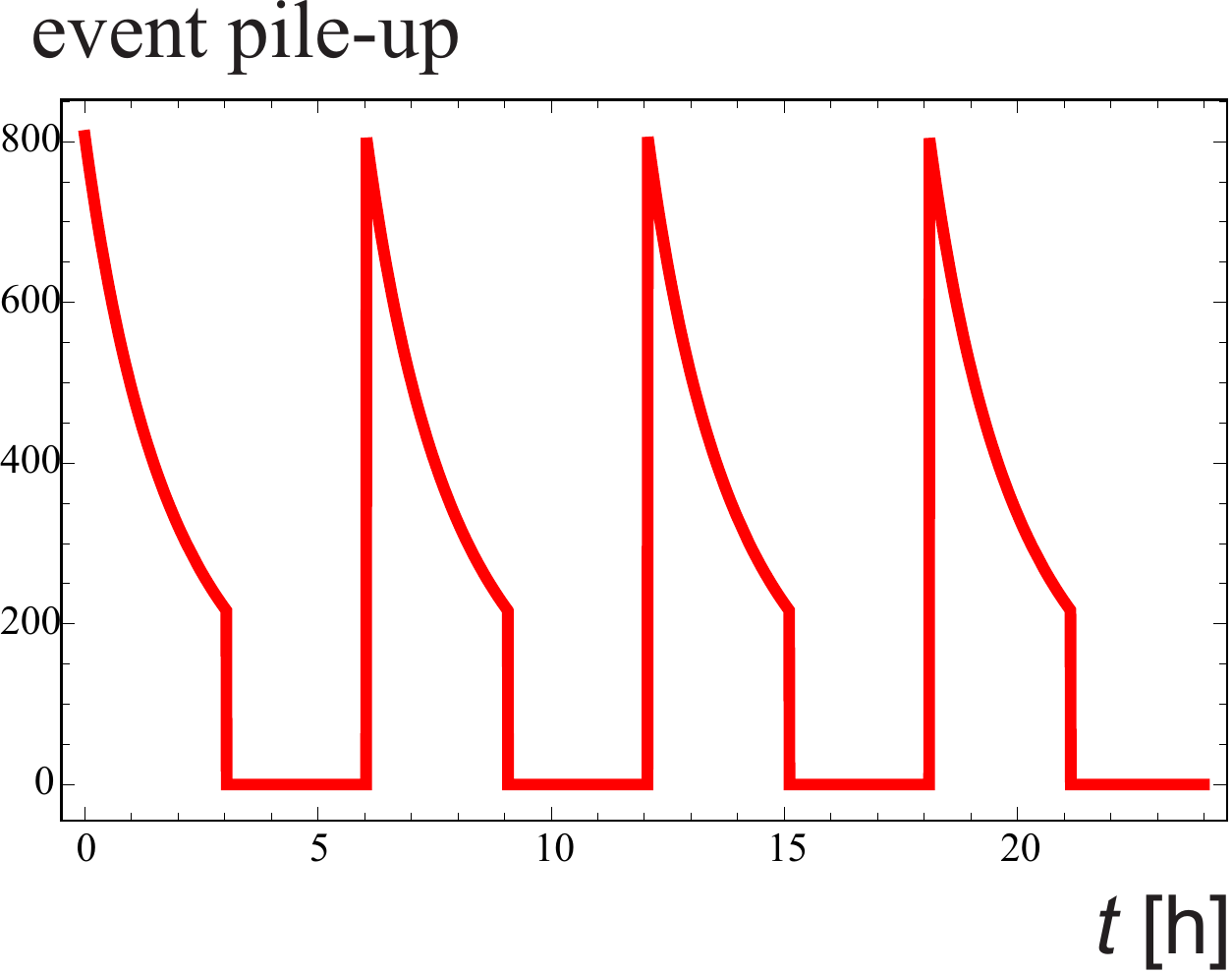}
\includegraphics[width=0.45\columnwidth]{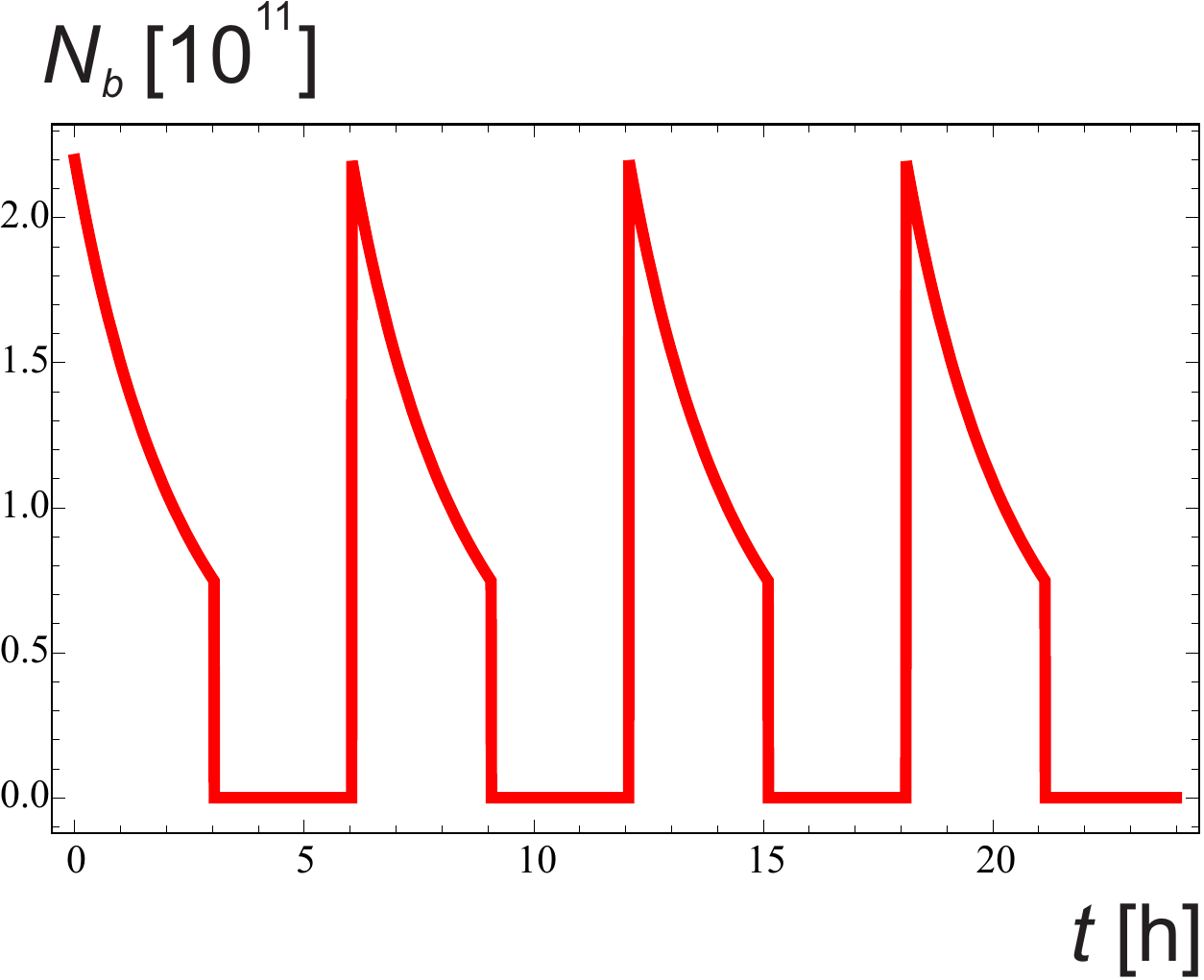}
\includegraphics[width=0.45\columnwidth]{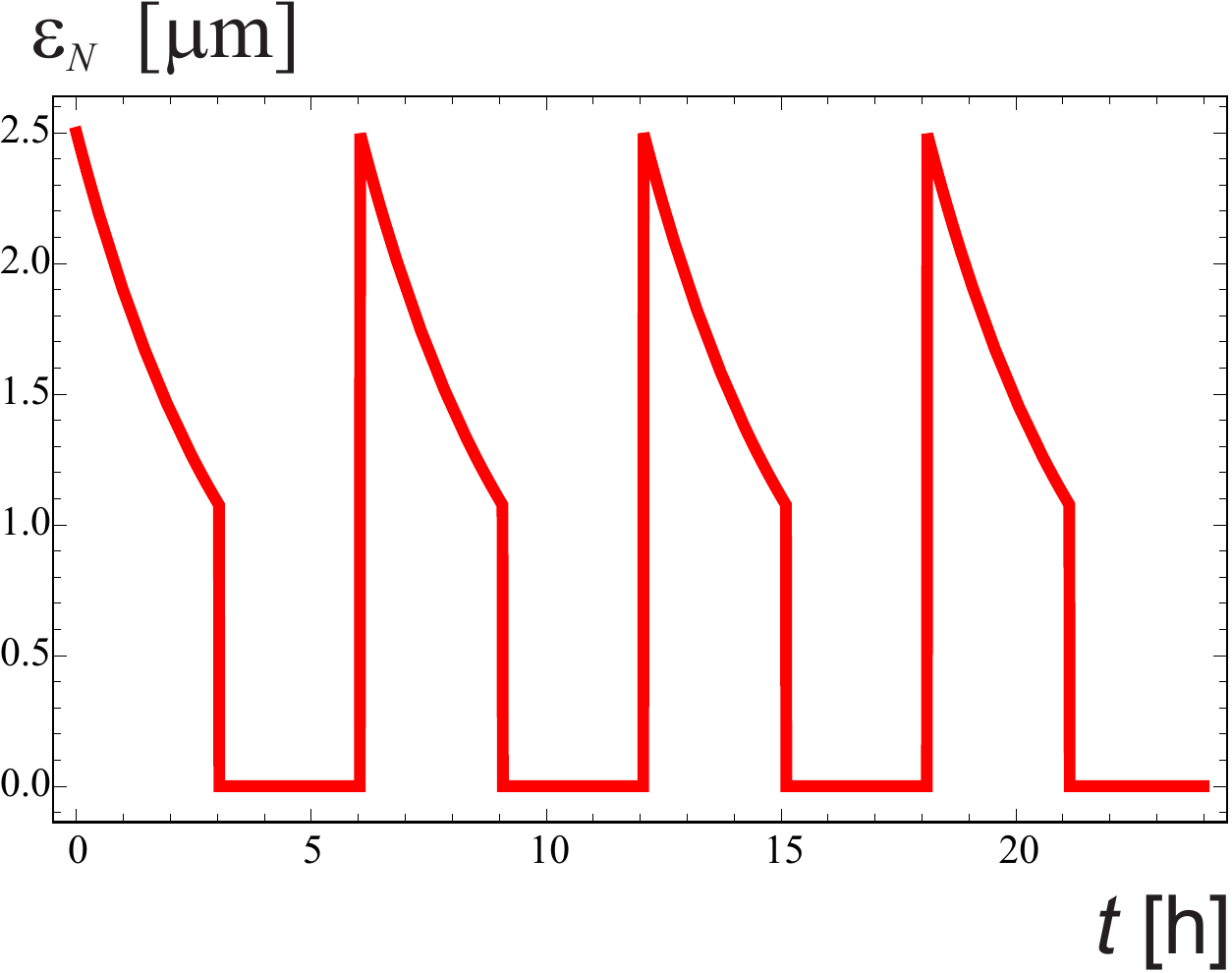}
\includegraphics[width=0.45\columnwidth]{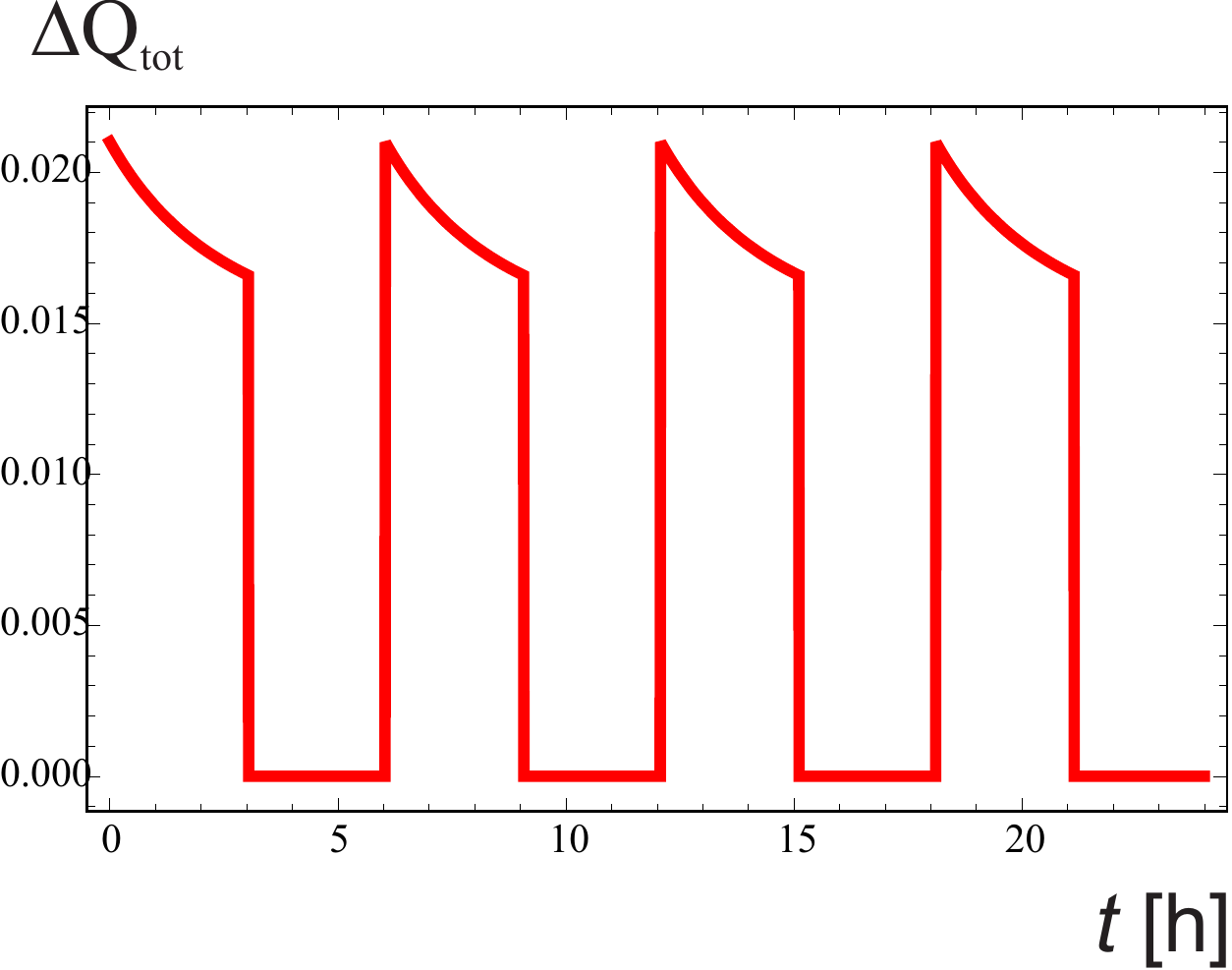}
\includegraphics[width=0.45\columnwidth]{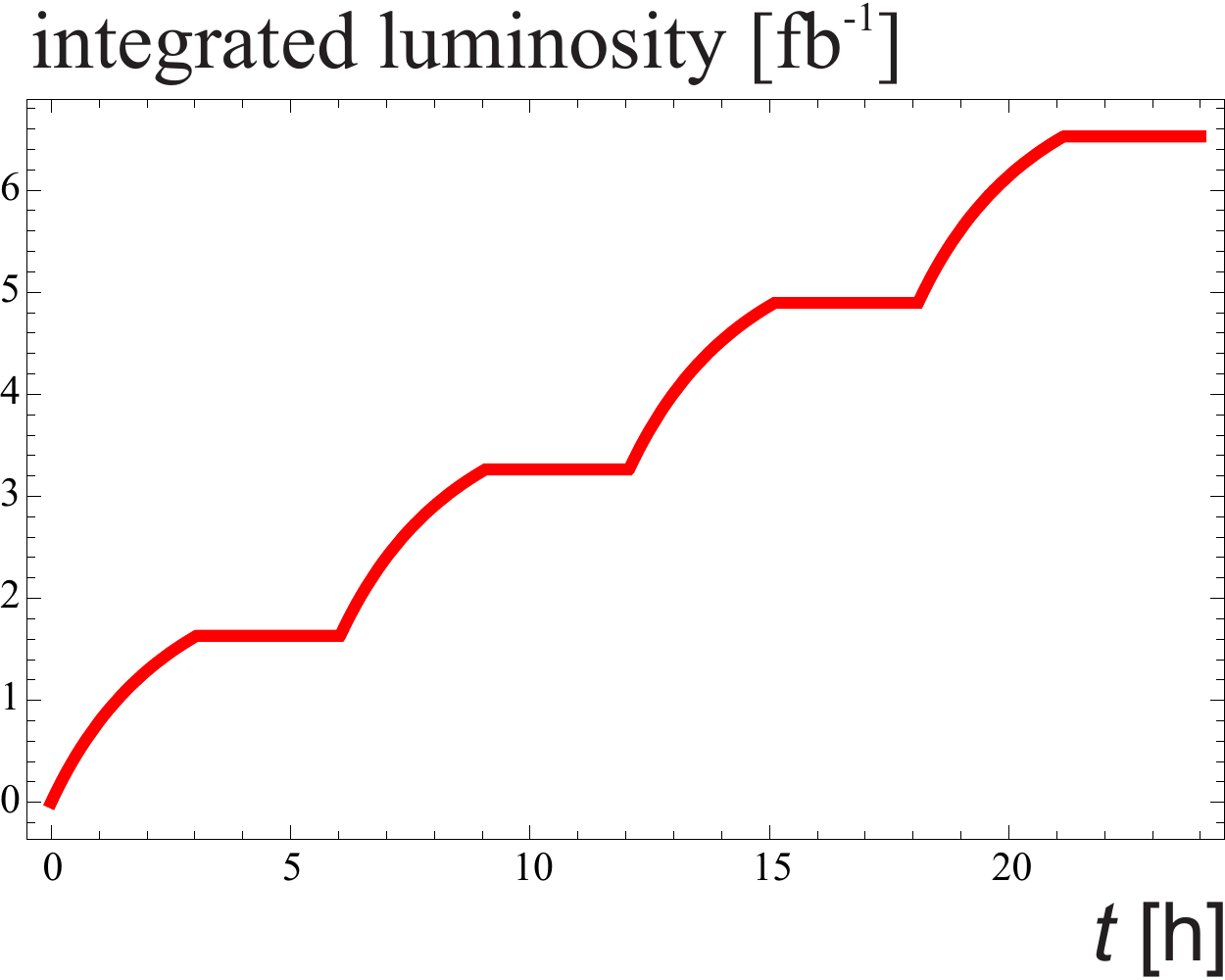}
\caption{Instantaneous luminosity, pile-up,
bunch population, normalized transverse 
emittance,  total beam-beam tune shift, 
and integrated luminosity 
as a function of time during 24 hours, for  
the HE-LHC at 100\%
machine availability. }
\label{fig:lumi-perf}
\end{figure}

Figure \ref{fig:lumi-perf} shows the time evolution 
of HE-LHC peak luminosity, pile-up, bunch intensity, transverse normalized emittance, total head-on beam-beam tune shift, and integrated luminosity over 24 hours at 100\% availability.

\begin{figure}[htp]
\centering
  \includegraphics[width=0.75\columnwidth]{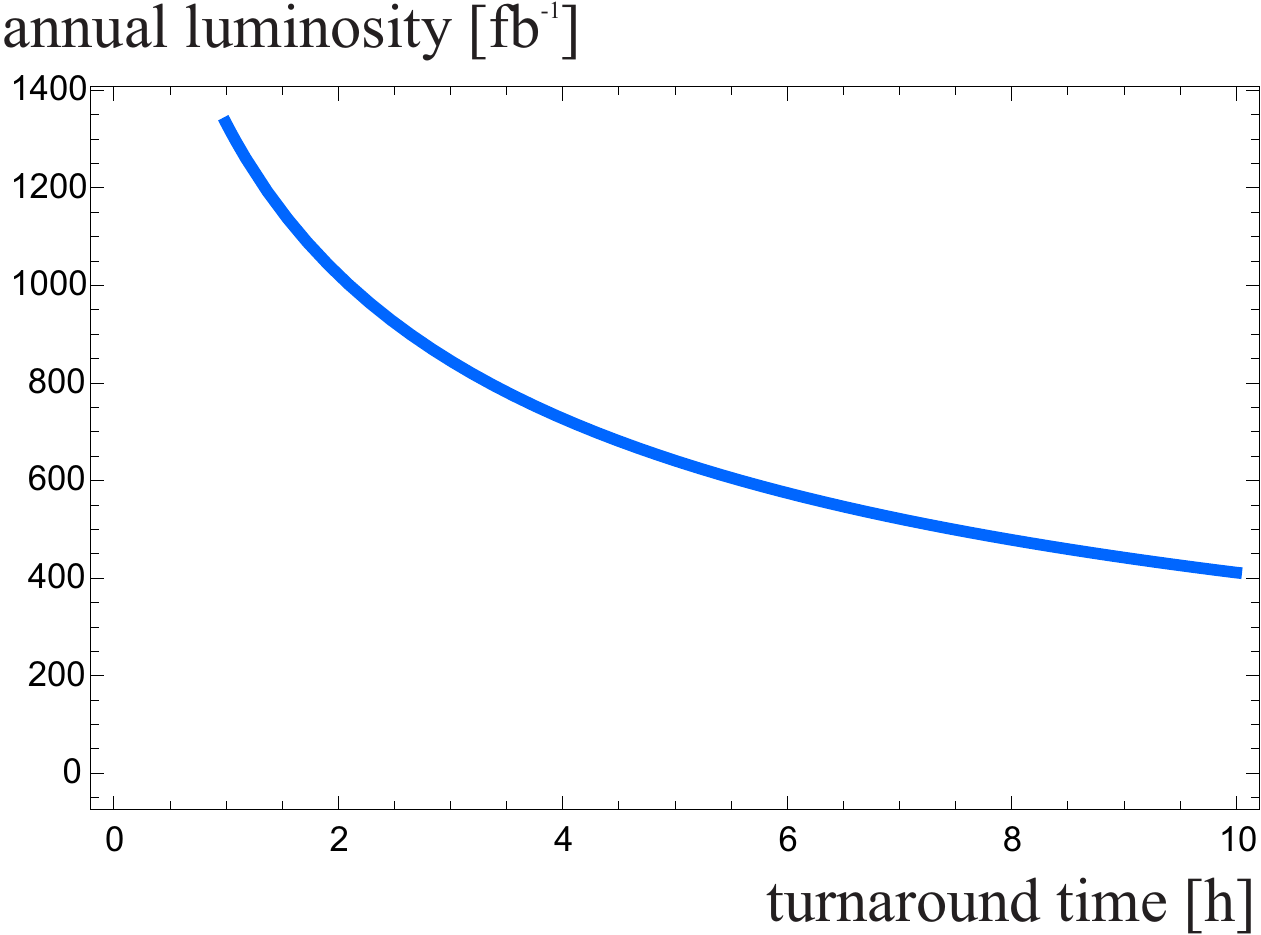}
\caption{Average annual luminosity versus average
turnaround time for the HE-LHC, assuming 70\% machine
availability and 160 days calendar days scheduled for physics operation per year. }
\label{fig:turnaround}
\end{figure}

The typical optimum run time of HE-LHC is about 3 hours.
The turnaround time should not be much longer than this. Otherwise the integrated luminosity performance significantly decreases, as is illustrated in Fig.~\ref{fig:turnaround}.

\section{Hadron-Collider Technologies}
\begin{figure}[htbp]
\centering
\includegraphics[width=0.9\columnwidth]{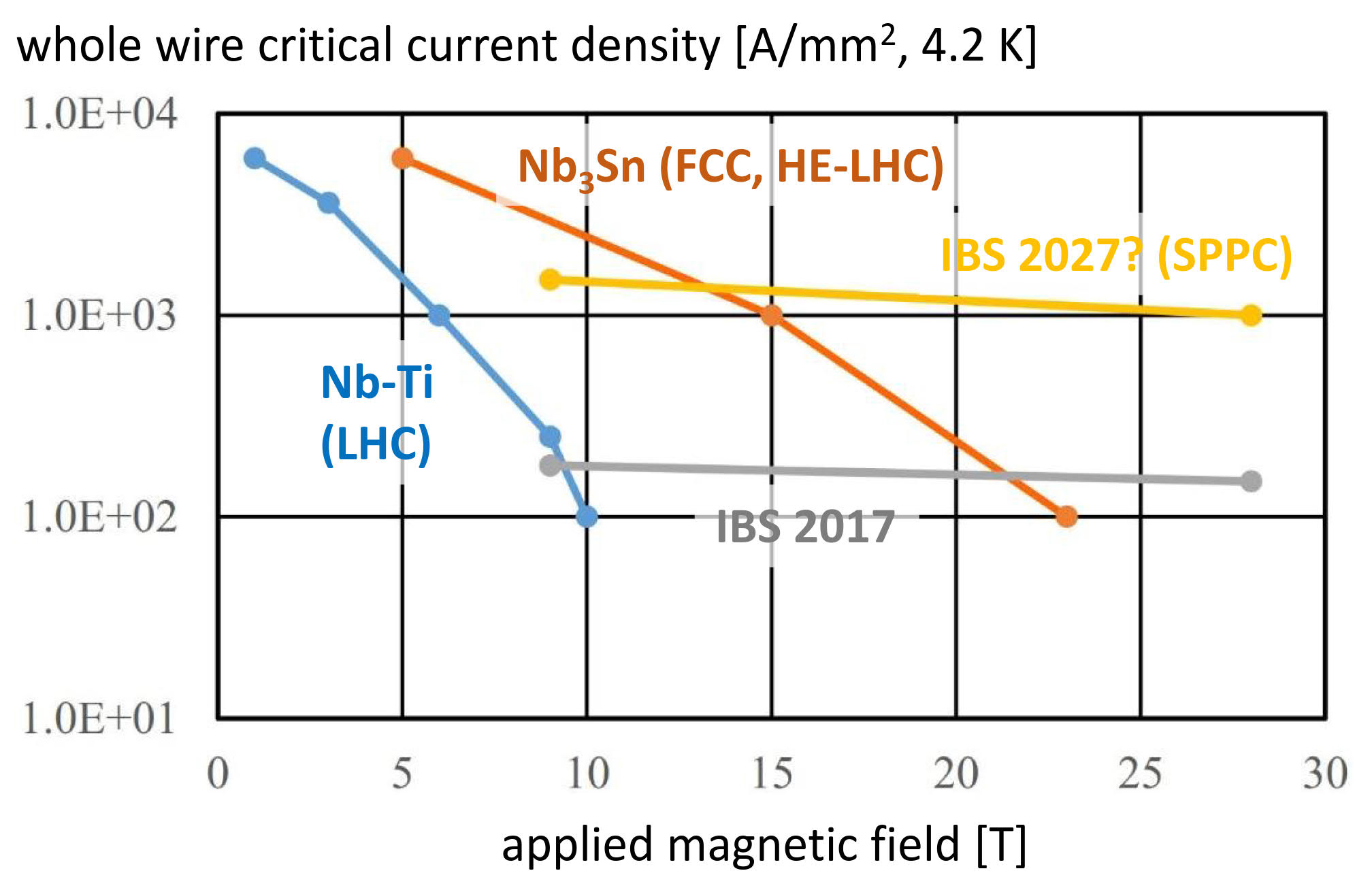}
\caption{Field limits for LHC-type Nb-Ti conductor, Nb$_{3}$Sn conductor as used for 
HL-LHC, FCC-hh and HE-LHC, and iron-based superconductor (IBS, present and 
10-year forecast) for SppC 
(after P.J.~Lee \protect\cite{plee}, and private 
communication J.~Gao).}
\label{sc}
\end{figure}

The primary technology of future hadron colliders is high-field magnets, both dipoles and quadrupoles. Magnets made from 
Nb-Ti superconductor were the core technology of the present LHC, Tevatron, RHIC and HERA. Nb-Ti magnets are limited to 
maximum fields of about 8 T.
 The HL-LHC will use, for the first time in a collider, some tens of dipole and quadrupole magnets with a peak field of 11-12 T, based on a new high-field magnet technology using Nb$_{3}$Sn superconductor. This will prepare the ground for the development of 16 Tesla Nb$_{3}$Sn magnets, and the later production of about 5000 Nb$_{3}$Sn magnets required by the FCC-hh.
The Chinese SppC magnets will utilize cables based on iron-based high-temperature superconductor (IBS), 
a material discovered at the Tokyo Institute of Technology in the year 2006 \cite{ibssc}.   
Figure \ref{sc} sketches the respective current densities and field limits. It is clear that 
Nb$_{3}$Sn can approximately double the magnetic field reached with Nb-Ti. 
The R\&D target for SppC looks agressive. 
The SppC goal is to increase the performance ten times while simultaneously reducing the cost by an order of magnitude.
If successful, the IBS magnet technology could become a game changer for future hadron colliders.

\begin{figure}[htbp]
\centering
\includegraphics[width=.65\linewidth]{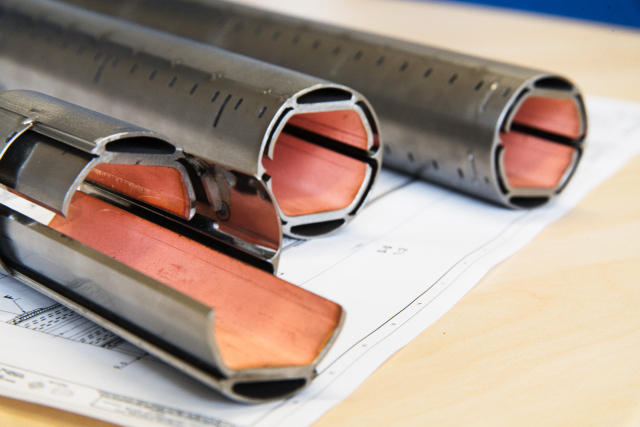}
\caption{\label{fcc-hh-bs}
Prototypes of the proposed FCC-hh vacuum chamber 
\protect\cite{kers3} (Image credit: CERN).}
\end{figure}

Another important technology is the cryo beam vacuum system, which has to cope with unusually high levels of synchrotron 
radiation in a cold environment, 
about 5 MW in total for FCC-hh. 
The design of the beam screen intercepting the radiation 
inside the cold bore of the magnets 
and the choice of its operating temperature (at 50 K significantly higher than the 5--20 K chosen for the LHC beam screen) are key ingredients of the 
hadron-collider design. 
First hardware prototypes for FCC-hh (see Fig.~\ref{fcc-hh-bs}) have been
tested with synchrotron radiation from an electron beam at the KIT ANKA facility at Karlsruhe in 2017.
These beam measurements have validated the basic 
design assumptions.

Further key technologies of the hadron collider include the collimators, the kicker and septa required for the extremely high
 beam energy, and the superconducting radiofrequency systems, e.g.~for acceleration and for compensation of synchrotron-radiation 
energy losses, as well as for the ever more demanding crab cavities.

\section{Hadron Collider Tunnel and Infrastructure}
\begin{figure}[htbp]
\centering
\includegraphics[width=0.98\columnwidth]{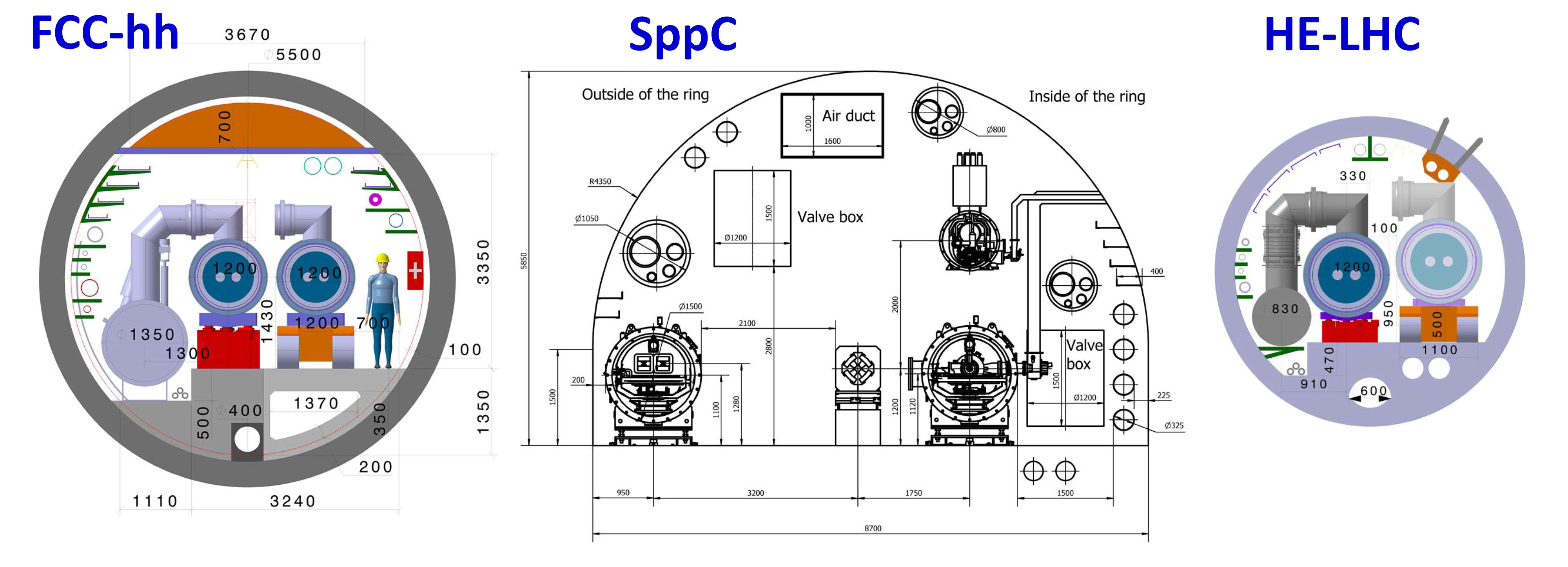}
\caption{Tunnel cross sections for HE-LHC, SppC and FCC-hh.}
\label{tunnel}
\end{figure}

The tunnel is a core element of any new collider.
The FCC-hh and SppC tunnels are constructed differently,
by tunnel boring machine or drill/blast 
technique, respectively.
The tunnel shapes and sizes are also rather different, as is illustrated in Fig.~\ref{tunnel}. 
The HE-LHC must fit into the existing LHC tunnel with a diameter of 3.8 m. Therefore, the HE-LHC dipole 
magnets must be made as compact as possible, with a maximum outer diameter of 1.2~m. 
In addition half-sector cooling is proposed to reduced the  diameter of the cryogenics lines and relax the tunnel integration, calling for additional 1.8 K refrigeration units. 
The new round tunnel for the FCC-hh will have a significantly larger diameter of 5.5 meter, to host the (possibly larger) 
16 T magnets and enlarged cryogenics lines, plus allow for additional safety features, such as smoke extraction, ventilation, escape passages etc. This large a tunnel still 
does not offer enough space to accommodate both a lepton and a hadron machine at the same time.
If in a first step the FCC-ee is built, it will need to be disassembled prior to the installation of 
the FCC-hh hadron collider.
The SppC tunnel is even larger, with a transverse width of 8.7 m. It is meant to provide enough space for both lepton 
and hadron machines, also including a lepton booster ring for top-up injection, which, in principle, could all be operated in parallel.

The HE-LHC will need eight new cryoplants, each with 28 kW equivalent cooling capacity at 4.5 K, e.g., about 1.5 times 
the capacity of one of the existing eight LHC plants, and additional plants at 1.8 K for the half sector cooling.   
In view of their much larger circumference and high synchrotron radiation power the FCC-hh and SppC will both need substantially larger cryogenic facilities still.
Specifically. the FCC-hh foresees 10 cryoplants, each with 50--100 kW at 4.5 K including
12 kW at 1.8 K, and requires a helium inventory of 800 tons, about 6 times the helium inventory of the
present LHC. 
The electrical power consumption of the FCC-hh cryoplants is about 200 MW \cite{laurent2017}.

\section{Time Lines and Cost}
\begin{figure}[htbp]
\centering
\includegraphics[width=0.98\columnwidth]{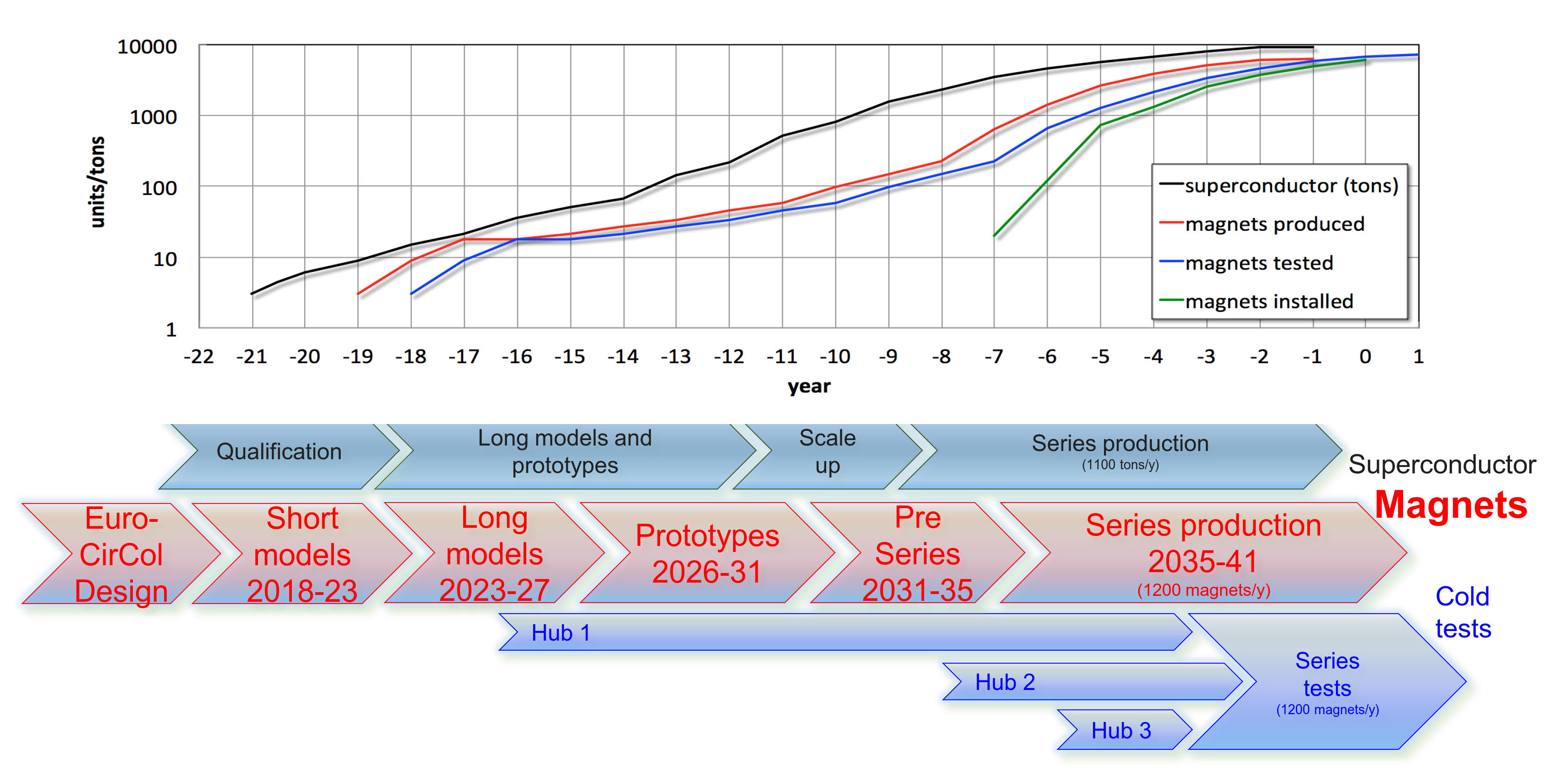}
\caption{Time line of FCC-16 T magnet R\&D.}
\label{magnet}
\end{figure}

\begin{figure}[htbp]
\centering
\includegraphics[width=0.98\columnwidth]{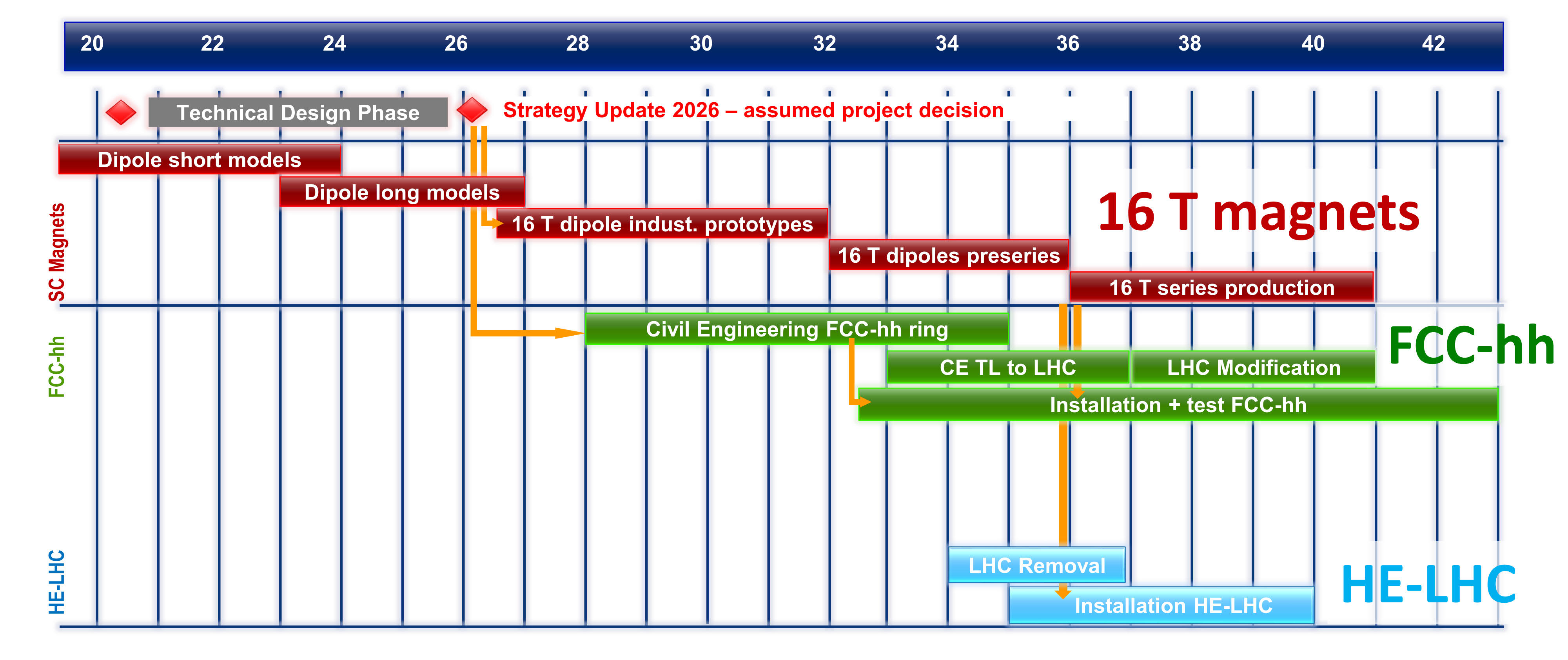}
\caption{Time line of FCC magnet production along with FCC-hh or HE-LHC implementation }
\label{implementations}
\end{figure}

The time line for FCC is determined by the time required for tunnel construction and by the magnet R\&D and production programme, as is illustrated in Figs.~\ref{magnet} and \ref{implementations}.
If HL-LHC stops in the Long Shutdown 5, presently 
scheduled around the year 2034, the HE-LHC could start physics operation in 2040.
FCC-hh would begin operation three years later, in 2043.
Very similarly, the latest time schedule for SppC foresees first hadron-beam collisions in 2045 \cite{gao2017}.

Construction cost for FCC-hh and SppC 
depends on the magnets, since several thousands of magnets are needed for each of these projects. 
So far the cost per magnet increases with the magnetic field and it is, in particular, higher for the novel magnet technologies. 
A milestone R\&D target for FCC is to make the price of an FCC 16 Tesla magnet the same as the price of an LHC 8.3 Tesla magnet.  
For SppC the R\&D goal is to increase the performance (current) per price unit of the IBS cable by two orders of magnitude over the coming 10 years.

In general, the construction cost of future projects can be 
minimized by \cite{shiltsev2,zimmermann2}:  
(1) reducing the cost of essential components,
in particular of the conductor material  
and the high-field magnets for the hadron colliders;
(2) building on a site with an existing infrastructure
and injector complex; and (3) staging, e.g., 
FCC-ee followed by FCC-hh, and, much later, possibly by a muon collider FCC-$\mu\mu$ \cite{zim-eaac}.

\section{Summary}
A future higher energy hadron collider will further push the energy frontier. Three such colliders are presently under study worldwide, with c.m.~energies ranging from 27 to 100 TeV, and perhaps even 150 TeV. 
R\&D on cost-effective high-field magnets is the key to their realization.  
Each of the three proposed colliders 
could start operation around the years 2040--2045.

\section*{Acknowledgements}
This report summarizes results from the international
FCC collaboration and from the CEPC/SPPC team. 
We are grateful to all collaboration members for their 
excellent work, and we are happy to acknowledge
specific contributions from V.~Mertens,  G.~Rumolo, L.~Tavian, 
and R.~Tomas.  
We most warmly thank W.~Barletta for inviting this article,
and M.~Giovannozzi for a careful reading of the manuscript.   
This work was supported, in part, 
by the European Commission under the HORIZON 2020 projects EuroCirCol, grant agreement 654305; EASITrain, grant agreement no. 764879; and ARIES, grant agreement 730871.

\bibliographystyle{model1-num-names} 
\bibliography{mybib}{}

\end{document}